\documentclass[%
superscriptaddress,
preprint,
footinbib,
amsmath,amssymb,
aps,
prl,
floatfix,
]{revtex4-1}

\usepackage{CJK}
\usepackage{graphicx}
\usepackage{dcolumn}
\usepackage{bm}
\usepackage{hyperref}
\usepackage[mathlines]{lineno}
\usepackage{natbib}
\usepackage{multirow}
\usepackage{boldline}
\usepackage{xcolor}
\usepackage{soul}

\begin{document}

\begin{CJK*}{UTF8}{}
\CJKfamily{gbsn}
\title{
Topological Determinants of Perturbation Spreading in Networks
}

\author{Xiaozhu Zhang (张潇竹)}
\email{xiaozhu.zhang@tu-dresden.de}
\affiliation{Chair for Network Dynamics, Institute for Theoretical Physics and Center for Advancing Electronics Dresden (cfaed), Cluster of Excellence Physics of Life, Technical University of Dresden, 01062 Dresden, Germany}
\author{Dirk Witthaut}
\affiliation{Institute for Energy and Climate Research - Systems Analysis and Technology Evaluation (IEK-STE), Forschungszentrum J\"ulich GmbH}
\author{Marc Timme}
\email{marc.timme@tu-dresden.de}
\affiliation{Chair for Network Dynamics, Institute for Theoretical Physics and Center for Advancing Electronics Dresden (cfaed), Cluster of Excellence Physics of Life, Technical University of Dresden, 01062 Dresden, Germany}
\affiliation{Department of Physics, Technical University of Darmstadt}

\date{\today}

\begin{abstract} 
Spreading phenomena essentially underlie the dynamics of various natural and technological networked systems, yet how spatiotemporal propagation patterns emerge from such networks remains largely unknown. Here we propose a novel approach that reveals universal features determining the spreading dynamics in diffusively coupled networks and disentangles them from factors that are system specific. In particular, we first analytically identify a purely topological factor encoding the interaction structure and strength, and second, numerically estimate a master function characterizing the universal scaling of the perturbation arrival times across topologically different networks. The proposed approach thereby provides intuitive insights into complex propagation patterns as well as accurate predictions for the perturbation arrival times. The approach readily generalizes to a wide range of networked systems with diffusive couplings and may contribute to assess the risks of transient influences of ubiquitous perturbations in real-world systems.

\end{abstract}


\maketitle

\end{CJK*}

Perturbation spreading centrally underlies many transient dynamics of networks, including in social, biological and infrastructural systems \cite{timme2019}. Examples include epidemic dynamics \cite{brockmann2013, iannelli2017, chen2018, moore2020, gautreau2007, karsai2011}, the signal propagation in biochemical networks \cite{maslov2007, maayan2005, santolini2018, hens2019, timme2004}, the spreading of information in social networks \cite{gernat2018, centola2010} as well as the impact of load shedding, infrastructure failures and fluctuations on the dynamics of power grids \cite{kettemann2016, tamrakar2018, haehne2019}. One main goal of studying perturbation spreading in networks is to predict at which time a perturbation at one network unit impacts other units \cite{iannelli2017, chen2018, moore2020, gautreau2007, hens2019, wolter2018}. Such times, among other consequences, influence whether and in which sense external perturbations may stay localized and tell how far stable network operation and control may be possible. For instance, in AC power grids where specific nodes are intrinsically exposed to fluctuating inputs, estimating the time taken for a contingency, e.g. a power surge caused by a squall hitting an off-shore wind farm, to cause the machine frequency exceeding the safety range at a distant node would provide valuable information about the time window for reactions and countermeasures.

Yet, disentangling the dominating factors in perturbation spreading dynamics remains a nontrivial task. A complex spatio-temporal spreading pattern results from the interplay of many contributions, including the underlying network topology, the state prior to perturbations, the units' intrinsic dynamics and the characteristics of the perturbation signal. For epidemic reaction-diffusion systems, the local dynamic response at each unit is stereotypical such that the spreading of an infectious disease is mainly determined by the effective distance between units \cite{brockmann2013, iannelli2017}. Gene regulatory networks and population dynamics also exhibit distinctive spreading patterns in response to changing static signals, respectively driven by graph distances and node degrees \cite{hens2019, timme2019}. Similarly, the spreading dynamics of AC power grids has been qualitatively categorized to be either ballistic or diffusive with respect to geometric distance depending on system inertia \cite{tamrakar2018}. A general understanding of the transient dynamics of spreading patterns is still missing.
 
Here we develop a theory of transient spreading dynamics that extracts the universal features underlying perturbation spreading patterns across topologically different networks. By analytically identifying a topological factor encoding the system's base state, we disentangle the topological determinants on transient spreading dynamics from other jointly influencing factors, and obtain a numerically estimated master function characterizing the universal scaling of the perturbation arrival times. We thus provide an intuitive understanding of the complex perturbation spreading patterns in diffusively coupled networks and propose an accurate prediction scheme for the perturbation arrival times.

\section{The topological factor in network transient responses}

\begin{figure}[t]
    \centering
    \includegraphics[width=0.5\textwidth]{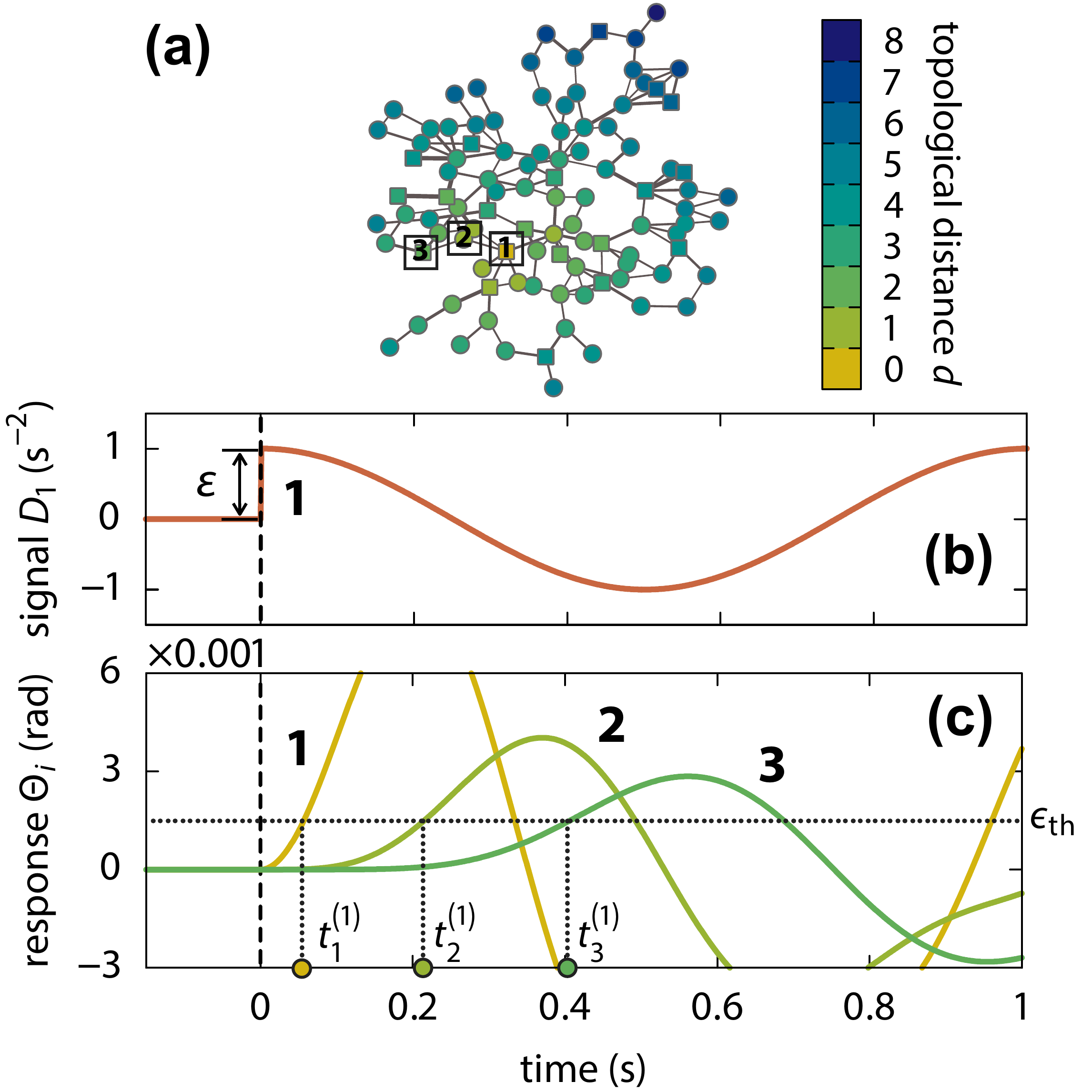}
    \caption{\textbf{Characterizing perturbation spreading in diffusively-coupled networks.} (a) Topology of a random network created based on \cite{schultz2014}. Each node is color-coded with its topological distance to the perturbed node 1. (b) Starting from $t=0$, node 1 is perturbed with a sinusoidal signal with frequency 1 Hz. (c) Response time series of three representative nodes $1,2,3$ at topological distances $d\in \{0,1,2\}$. The times $t_1^{(1)},t_2^{(1)},t_3^{(1)}$ at which the responses at these nodes cross a threshold $\epsilon_{\mathrm{th}}=1.5\times10^{-3}\;\mathrm{rad}$ are illustrated. The network parameters are set as $K_{ij}=30\;\text{s}^{-2}$, $\alpha=1\;\text{s}^{-1}$, $\omega_+=4\;\text{s}^{-2} $ [squares in (a)], $\omega_-=-1\;\text{s}^{-2}$ [discs in (a)]. Each node is assigned with a natural frequency of $\omega_+$ or $\omega_-$ at random, satisfying $\sum_i\omega_i=0$ so that a fixed point exists for the given network topology.}
    \label{fig:illustration}
\end{figure}

We here focus on networks of $N$ second-order Kuramoto oscillators, whose dynamics is governed by the equations of motion
\begin{equation}
    \ddot{\theta}_i=\omega_i-\alpha\dot{\theta}_i+\sum_{j=1}^NK_{ij}\sin(\theta_j-\theta_i)+D_i^{(k)}(t)
\end{equation}
with $i\in\{1,\cdots,N\}$. Here $\theta_i$ is the phase angle of oscillator $i$, $\omega_i$ is the natural acceleration of oscillator $i$, and $\alpha$ and $K_{ij}$ parametrize the system damping and the coupling strength between oscillator $i$ and $j$ respectively. The particular model characterizes the dynamics of AC power transmission networks, where the synchronous machines with damping $\alpha$ operate with angles $\theta_i$ relative to the reference frame rotating at the nominal grid frequency $2\pi\times 50$ Hz \footnote{In the US and parts of Japan, the nominal grid frequency is 60 Hz.} and are connected by transmission lines with coupling strength $K_{ij}$. We study the collective transient responses in a stable operational state $\bm{\theta}^*=(\theta^*_1,\cdots,\theta^*_N)^\mathsf{T}$ to an external perturbation to the natural acceleration at oscillator $k$, reflecting the change in power production ($\omega_k=\omega_+>0$) or consumption ($\omega_k=\omega_-<0$) in power grids. The sinusoidal perturbation at node $k$ starting at $t=0$ is represented by an $N$-dimensional perturbation vector $\bm{D}^{(k)}(t)$ with only one non-zero element, i.e. $D_i^{(k)}(t)=H(t)\delta_{ik}\varepsilon e^{\imath(\Omega t+\varphi)}$ with $H(\cdot)$ being the Heaviside step function (see Fig.~\ref{fig:illustration}) \footnote{The network response to generic perturbation is given by a superposition across input nodes and frequencies \cite{zhang2019}}. If the perturbation is not too strong to desynchronize the system  or to cause a line overload, that is, $|\theta_j-\theta_i| \leq \pi/2$ \footnote{Given that at the fixed point $|\theta^*_j-\theta^*_i|<\pi/2$ is satisfied for all edges $(i,j)$ \cite{manik2014}.}, the $N$-dimensional trajectory of the network response is accurately provided by a linear response theory \cite{zhang2019}, that is, the solution of the linearized dynamics
\begin{equation}
    \ddot{\mathbf{\Theta}}^{(k)}=-\alpha\dot{\mathbf{\Theta}}^{(k)}-\mathcal{L}\mathbf{\Theta}^{(k)}+\bm{D}^{(k)}(t)
    \label{eq:linear}
\end{equation}
for the response vector $\bm{\Theta}^{(k)}(t):=\bm{\theta}(t)-\bm{\theta}^*$. Here $\mathcal{L}$ denotes a weighted graph Laplacian matrix depending on the network's base state: $\mathcal{L}_{ij}:=-K_{ij}\cos(\theta^*_j-\theta^*_i)$ for $i\neq j$ and $\mathcal{L}_{ii}:=-\sum_{j\neq i}\mathcal{L}_{ij}$.  

The equation \eqref{eq:linear} admits a formal analytic solution (see Supplemental Material Sec.~S1), but such a solution does not reveal which are the essential factors that determine the complex spatio-temporal pattern of perturbation spreading. 
To focus on the transient spreading dynamics, we extract the transient parts from the linear response solution by representing it as a power series of time close to $t=0$. We observe the matrix of linear responses, $\bm{\varTheta}$, with its $(k,i)$-element being the linear response $\Theta_i^{(k)}(t)$ at unit $i$ to a single perturbation at unit $k$ at a graph-theoretic (topological) distance $d=d(k,i)$. The $(2d+2)$-th order term in the power series, which contains the $(2d+2)$-th time derivative $\hat{D}_t^{2d+2}\bm{\varTheta}^{(k)}(t)$ \footnote{$\hat{D}_t^{n}:=\frac{d^n}{dt^n}$ is the differential operator, indicating the action of taking the $n$-th order derivative with respect to time $t$, not to be confused with $D^{(k)}_i(t)$, the perturbation signal at node $i$.} of the linear response matrix $\bm{\varTheta}$, is a polynomial of the weighted graph Laplacian $\mathcal{L}$ with degree $d$  (see Supplemental Material Sec.~S2 for a proof). For linear response $\Theta_i^{(k)}(t)$, the terms with an order lower than $2d+2$ in the power series vanish, leading to the transient response
\begin{equation}
    \Theta_i^{(k)}(t)
    =\dfrac{\hat{D}_t^{2d+2}\Theta_i^{(k)}(0)}{\left(2d+2\right)!}t^{2d+2}+O\left(t^{2d+3}\right)
	 \label{eq:series}
\end{equation}
asymptotically as $t\rightarrow 0$, because these lower order terms contain the $(k,i)$-element of $\mathcal{L}^p$ with $p<d$, which are exactly zero. For the same reason, the $(2d+2)$-th order term contains only one non-zero term, i.e. the highest order term proportional to $\left(\mathcal{L}^d\right)_{ki}$.
This leading-term approximation of the transient response enables to estimate the arrival time $t_i^{(k)}$, i.e. the time when the response \footnote{Here the response, as a physical quantity, refers to the real part of the complex extension of $\Theta_i^{(k)}(t)$.} exceeds a given threshold, $\Theta_i^{(k)}(t)\geq \epsilon_{\mathrm{th}}$, [illustrated in Fig.~\ref{fig:illustration}(c)] as
\begin{equation}
    t^{(k)}_{i,\text{est}}
    =\left|\dfrac{\epsilon_{\text{th}}(2d+2)!}
    {\varepsilon \cos(\varphi)\left(\mathcal{L}^{d}\right)_{ki}}
    \right|^{\frac{1}{2d+2}}.
	 \label{eq:tarr}
\end{equation}
The estimated arrival time \eqref{eq:tarr} reveals that the impact of network topology on the transient response dynamics is concentrated in   
\begin{equation}
    T_{i}^{(k)}:=\left|\left(\mathcal{L}^{d}\right)_{ki}\right|^{-\frac{1}{2d+2}},
    \label{eq:topofactor}
\end{equation}
which we call the \textit{topological factor} in the perturbation spreading process from $k$ to $i$. The topological factor $T_{i}^{(k)}$ encodes all information about the specific network setting prior to perturbations: the network structure, the coupling strengths and the network flow pattern at the steady state. The other factor in \eqref{eq:tarr},
\begin{equation}
    \dfrac{t^{(k)}_{i,\text{est}}}{T_i^{(k)}}=\left[C(2d+2)!\right]^{\frac{1}{2d+2}},
    \label{eq:distfactor}
\end{equation}
characterizes a generic spreading pattern in diffusively-coupled networks. Here $C=\left|\epsilon_{\text{th}}/\varepsilon \cos(\varphi)\right|$ is a network-independent constant quantifying the relative response level, i.e. the ratio between the response threshold and the initial perturbation magnitude near $t=0$.

\begin{figure*}
    \centering
    \includegraphics[width=\textwidth]{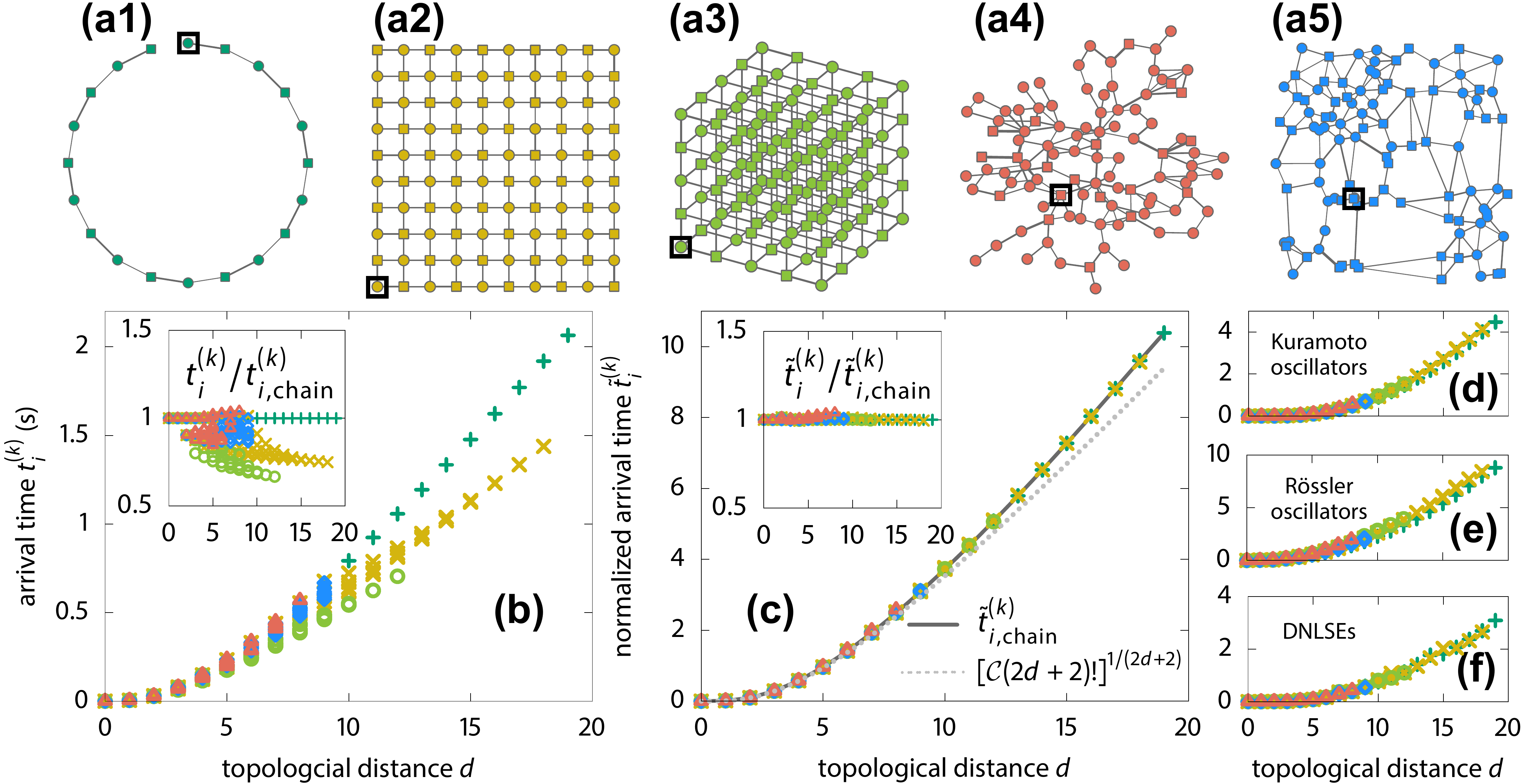}
    \caption{\textbf{Universal spreading dynamics across diverse network topologies.} (a) Five diverse network topologies: chain (a1), square lattice (a2), cubic lattice (a3), random power grid \cite{schultz2014} (a4) and Delaunay triangulation networks (a5). The perturbed node is marked with a black square. (b) The threshold-crossing arrival times ($\epsilon_{\text{th}}=10^{-9}$ rad) from direct simulations are scattered, however (c) they largely collapse across topologies if normalized by the respective topological factors, $\tilde{t}_i^{(k)}:={t}_i^{(k)}/T_i^{(k)}$. (b-f) share the color-coding with the network sketches (a1)-(a5). The arrival time differences among network topologies, e.g. the ratio between the arrival times with topological distance $d$ and the ones with the same $d$ in the chain network are highlighted in the insets of (b) and (c). The dotted line in (c) illustrates the topology-independent factor \eqref{eq:distfactor}, explicating the differences from the normalized actual arrival times (solid grey line as a visual aid). (d-e) In networks of (1-dimensional) Kuramoto oscillators (d), in networks of (3-dimensional) R\"ossler oscillators (e) and in networks of discrete nonlinear Schr\"odinger equations (DNLSEs) (f), normalized arrival times all collapse. For (a-c), the parameters for the networks ($\omega_i,K_{ij},\alpha$) and for the perturbation signal ($\varepsilon,\Omega,\varphi$) are the same as in Fig.~\ref{fig:illustration}, except for (a1) - (a3) $\omega_+=1\text{s}^{-2}, \omega_-=-1\text{s}^{-2}$ are distributed alternatively. For (e) R\"ossler oscillators ($a=0.1,b=0.1,c=4$) are diffusively coupled in variable $z$ and for (f) diffusive couplings naturally arise in the equation of motion for the complex field $\psi$ (see Supplemental Material Sec.~S3 for simulation details).}
    \label{fig:topofactor}
\end{figure*}

Direct simulations confirm that the topological factor $T_{i}^{(k)}$ \eqref{eq:topofactor} essentially captures the network effect on perturbation spreading dynamics. The arrival times scattered for diverse network topologies largely collapse once normalized by the topological factor, independent of the dimensionality of embedding and the topological irregularities [see Fig.~\ref{fig:topofactor}]. 
The topological factor can be intuitively interpreted as the inverse of an average of the \textit{edge spreading strengths} \footnote{The spreading strength of an edge $(u,v)$ corresponds to  $\left|\mathcal{L}_{uv}\right|$, which characterizes how strongly the initial flow reacts to the related phase changes.} along all shortest paths, reflecting how strongly two nodes are connected in terms of spreading. We further highlight that, in contrast to epidemic spreading dynamics \cite{brockmann2013, iannelli2017}, the perturbation arrival times in diffusively-coupled networks are \textit{non-additive} in distance, which essentially reflects the distance-dependency of local response activities, e.g. a decaying response slope for a growing distance [see Fig.~\ref{fig:illustration}(c)] (see Supplemental Material Sec.~S8 for a detailed discussion).

\section{Prediction of perturbation arrival times}

While the leading-term estimation of transient network responses \eqref{eq:tarr} succeeds in analytically uncovering the role of network topology in perturbation spreading dynamics (Fig.~\ref{fig:topofactor}), its value increasingly deviates from the actual arrival times as a perturbation spreads further [up to  about $10\%$, compare the solid and the dashed line in Fig.~\ref{fig:topofactor}(c)]. Such deviations originate from the neglected higher order terms in the power series \eqref{eq:series}. How can we estimate the contribution of the neglected terms which enables quantitatively accurate predictions for threshold-crossing arrival times?

We propose a semi-analytical prediction scheme based on the estimation of a master function, which characterizes the spreading dynamics and is largely independent of the underlying network topology. To accurately estimate the relative contribution of the higher order terms in transient responses \eqref{eq:series}, we numerically sample a master function
\begin{equation}
    M(d):=\left\langle t^{(k)}_{i}/T^{(k)}_{i}\right\rangle_d
    \label{eq:master}
\end{equation} 
by averaging the normalized actual arrival times $t^{(k)}_{i}/T_i^{(k)}$ at a specific distance $d$ for an ensemble of network topologies (``training data set''). We predict the arrival times for networks with topologies in the same ensemble but not in the training set by combining the master function and the specific topology in the target network. Particularly, we multiply the topological factor $T_{i'}^{(k')}$ corresponding to the spreading process from node $k'$ to node $i'$ in the target network by the master function at distance $d'=d(k',i')$:
\begin{equation}
    t^{(k')}_{i',\mathrm{pred}}=M(d')T_{i'}^{(k')}.
\end{equation}
Furthermore, we find a characteristic prediction interval $\Delta t^{(k')}_{i',\mathrm{pred}}:=h\sigma(d')T_{i'}^{(k')}$ given in terms of $\sigma(d)$, the standard deviation of the normalized arrival times for distance $d$ in the training data set. Here $h$ is a parameter tuning the width of the prediction interval.

\begin{figure}[h]
    \centering
    \includegraphics[width=0.5\textwidth]{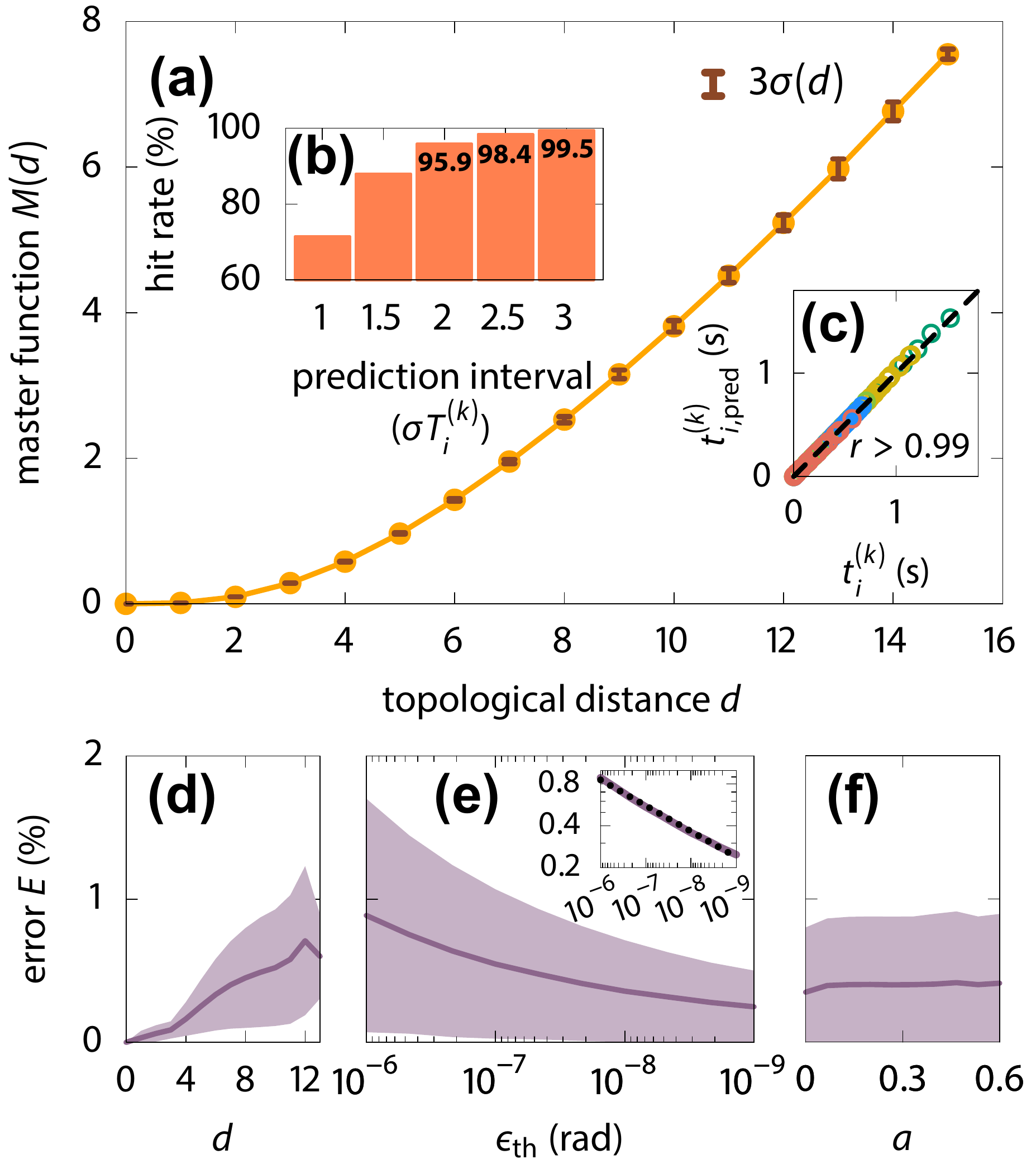}
    \caption{\textbf{Master function of spreading dynamics and prediction performance.} (a) A master function \eqref{eq:master} estimated from $10^6$ arrival time measurements in $100$ random network topologies \cite{schultz2014} with $100$ nodes. The error bars indicate a prediction interval of $3\sigma(d)$. (b) The arrival time predictions for the $5$ example networks in Fig.~\ref{fig:topofactor} highly correlate with the measurements (Pearson's correlation coefficient $r>0.99$). (c) In the prediction of $10^6$ arrival times in another $100$ random networks \cite{schultz2014}, the hit rate, i.e. the probability that the actual arrival time from simulation falls in the prediction interval, grows rapidly with a broadening prediction interval. (d-e) The relative prediction error $E:=|t^{(k)}_{i,\mathrm{pred}}-t^{(k)}_{i}|\big/t^{(k)}_{i}$ drops for smaller arrival times, e.g. (d) for smaller distances and (e) for lower thresholds. For a lowering threshold, the error decays approximately as a power law [inset of (e)]. (f) The prediction performance is robust against an increasing heterogeneity $a$ in network coupling strength. In (d-f) the shaded area indicates standard deviations. For (a-d,f) the threshold $\epsilon_{\mathrm{th}}=10^{-9}$ rad. For (a-e) the parameters ($\alpha$, $\omega_i$, $\varepsilon$, $\Omega$, $\varphi$, $K_{ij}$) are set the same as in Fig.~\ref{fig:illustration}. }
    \label{fig:prediction}
\end{figure}

The master function approach accurately describes the arrival dynamics of the spatio-temporal spreading patterns in networks with diffusive couplings. Similar semi-analytical studies for networks of other diffusively coupled units, including first-order phase oscillator networks, describing neural oscillations \cite{cumin2007,breakspear2010} and coupled Josephson junctions \cite{wiesenfeld1998}, networks of more complex, three-dimensional units such as R\"ossler oscillators, and non-oscillatory systems of coupled discrete nonlinear Schr\"odinger equations (DNLSEs), indicate qualitatively the same universal scaling featuring \eqref{eq:distfactor}, with a suitable adapted distance dependence [see Fig.~\ref{fig:topofactor}(d)-\ref{fig:topofactor}(e) and Supplemental Material Sec.~S3].
The prediction performance is robustly high (with the relative error below $1\%$) for an increasing heterogeneity in coupling strengths [see Fig.~\ref{fig:prediction}(f) and \footnote{For a fixed network topology the coupling heterogeneity is tuned by randomly drawing $K_{ij}$ from the uniform distribution on $[K(1-a/2),K(1+a/2)]$ with $K=30\text{ s}^{-2}$. Here $a$ defines the relative coupling spread between the maximal and minimal possible coupling strength, thus quantifying the network heterogeneity. For each $a$, the error is averaged over ten realizations.} for simulation details], and mildly drops as the arrival time grows \footnote{Because the master function relies on the approximated responses in the asymptotic regime as $t \rightarrow 0$, the contribution of the higher order terms grows and is harder to estimate as the response grows.}, e.g. with a larger distance and a higher response threshold [see  Fig.~\ref{fig:prediction}(d)-\ref{fig:prediction}(e)]. Multiple factors such as the characteristic degree distribution, the perturbation frequency, and the system dissipation rate also influence the specific shape of the master function, but not its predictive power in the respective system class (see Supplemental Material Sec.~S4-S6 for details).

We have thus disentangled the impact of network topology, including coupling structure and strength, from other factors jointly influencing the transient spreading dynamics in diffusively coupled networks, enabling us to accurately predict perturbation arrival times. Going beyond a formal linear response theory and series expansion, we identified the factor that captures the influence of network topology on transient perturbation spreading. It also reveals a high directional symmetry in perturbation spreading, i.e. $T_i^{(k)}= T_k^{(i)}$, meaning that the perturbation spreading between a node pair takes almost the same amount of time in both directions, independent of which node is perturbed even when one of them is a high-degree hub.
Beyond the topological factor, a master function for transient spreading dynamics estimates the relative contribution of higher order terms neglected in the asymptotic analysis and enables accurate predictions of arrival times. Local structural changes in large networks, such as upgrading units or integrating a new wind farm or consumer, may significantly alter the spreading pattern while the network degree distribution is hardly affected. The master function approach would be a powerful solution here: it still provides accurate predictions yet saves substantial computation time and power that would otherwise be required for repeated simulations needed by direct numerical analysis.

Our approach readily generalizes to the transient (spreading) dynamics in a wide range of networked systems with diffusive couplings, where a weighted graph Laplacian naturally arises in the network's linear response dynamics \eqref{eq:linear}, for instance networks with unit dynamics of different orders, different perturbations and different observables (see Supplemental Material Sec.~S3).
	
\begin{acknowledgments}
This project is funded by the Deutsche Forschungsgemeinschaft (DFG; German Research Foundation) under Germany's Excellence Strategy EXC-2068-390729961-Cluster of Excellence Physics of Life and the Center for Advancing Electronics at TU Dresden, by the German Federal Ministry for Research and Education (BMBF grants no. 03SF0472F and 03EK3055), and by the Helmholtz Association (via the joint initiative ``Energy System 2050-A Contribution of the Research Field Energy" and grant no. VH-NG-1025). 
\end{acknowledgments}

\bibliographystyle{apsrev4-1}
\nocite{zhang2018}
\bibliography{transient}

\end{document}